\documentstyle[aps]{revtex}
\def\ut#1{\rlap{\lower1ex\hbox{$\sim$}}#1{}}
\begin{document}

\draft
\title{Bosonic string theory with constraints linear in the momenta}

\author{Merced Montesinos \footnote{E-mail: merced@fis.cinvestav.mx}}

\address{Departamento de F\'{\i}sica, Centro de Investigaci\'on y de
Estudios Avanzados del I.P.N.,\\
Av. I.P.N. No. 2508, 07000 Ciudad de M\'exico, M\'exico.}

\author{Jos\'e David Vergara\footnote{E-mail: vergara@nuclecu.unam.mx}}

\address{Instituto de Ciencias Nucleares, Universidad Nacional
Aut\'onoma de M\'exico, 70-543, Ciudad de M\'exico, M\'exico}

\date{\today}
\maketitle
\begin{abstract}
The Hamiltonian analysis of Polyakov action is reviewed putting emphasis in
two topics: Dirac observables and gauge conditions. In the case of the closed
string it is computed the change of its action induced by the gauge
transformation coming from the first class constraints. As expected, the
Hamiltonian action is not gauge invariant due to the Hamiltonian constraint
quadratic in the momenta. However, it is possible to add a boundary term to
the original action to build a fully gauge-invariant action at first order.
In addition, two relatives of string theory whose actions are fully
gauge-invariant under the gauge symmetry involved when the spatial slice is
closed are built. The first one is pure diffeomorphism in the sense it has no
Hamiltonian constraint and thus bosonic string theory becomes a sub-sector of
its space of solutions. The second one is associated with the tensionless
bosonic string, its boundary term induces a canonical transformation and the
fully gauge-invariant action written in terms of the new canonical variables
becomes linear in the momenta.
\end{abstract}


\section{Canonical analysis}
Relativistic free strings propagating in an arbitrary $D$-dimensional fixed
background spacetime with metric $g=g_{\mu\nu}(X) dX^{\mu} dX^{\nu} $; $\mu,
\nu =0,1,...,D-1$, can be described, for instance, with the Polyakov action
\cite{Polyakov}
\begin{eqnarray}
S[\gamma^{ab}, X^{\mu}] & = &
\alpha \int_M d ^2\xi \sqrt{-\gamma} \gamma^{ab} \partial_a X^{\mu}
\partial_b X^{\nu}  g_{\mu\nu} (X) \label{polyakov}\, .
\end{eqnarray}
The variation of $S[\gamma^{ab}, X^{\mu}]$ with respect to the background
coordinates $X^{\mu}$ and the metric $\gamma^{ab}$ yields the equations of
motion
\begin{eqnarray}
\nabla^{a} \nabla_{a} X^{\mu}+ \Gamma^{\mu}\,_{\theta\phi} \,\, \gamma^{bc}
\partial_b X^{\theta} \partial_c X^{\phi}& = & 0 \, , \nonumber\\
T_{ab} := \frac{\alpha}{2} \gamma_{ab} \gamma^{cd}
\partial_c X^{\mu} \partial_d X^{\nu} g_{\mu\nu} -
\alpha \partial_a X^{\mu} \partial_b X^{\nu} g_{\mu\nu} & = &
\frac{\alpha}{2} \gamma_{ab} \gamma^{cd}
h_{cd} - \alpha h_{ab} = 0 \, , \label{tensor}
\end{eqnarray}
respectively. Here, $h_{ab} = \partial_a X^{\mu} \partial_b X^{\nu}
g_{\mu\nu}$ is the induced metric on the world sheet, $\nabla$ is the
covariant derivative associated with the Levi-Civita connection of
$\gamma_{ab}$,  $\Gamma^{\mu}\,_{\theta\phi}$ are the Christoffel symbols
associated with the background metric $g_{\mu\nu}$. The Lagrangian formalism
is more common than the Hamiltonian one for string people community. However,
the Hamiltonian framework is a necessary step to perform the quantization of
the theory using Dirac's method \cite{Dirac}. Also the Hamiltonian framework
is the natural arena to analyze the issues of observables and gauge
conditions for the theory which have relevance both in its classical and
quantum dynamics. That is why here the canonical analysis is reviewed putting
emphasis in these two topics. To go to the Hamiltonian formalism, it is
mandatory to choose a time coordinate $\xi^0=\tau$ and a space coordinate
$\xi^1=\sigma$ and assume that the world sheet $M$ has the topology $M=
R\times \Sigma$. The metric $\gamma_{ab}$ is put in the ADM form
\begin{eqnarray}
\left ( \gamma_{ab} \right ) =  \left (
\begin{array}{ll}
  - N^2 + \lambda^2  \chi & \chi \lambda \\
\,\,\,\,\,\chi \lambda & \chi
\end{array}
\right )\, ,  \quad
\left ( \gamma^{ab} \right ) =  \left (
\begin{array}{ll}
-\frac{1}{N^2} & \frac{\lambda}{N^2}\\
\,\,\,\,\,\frac{\lambda}{N^2} & \frac{1}{\chi} - \frac{\lambda^2}{N^2}
\end{array}
\right )\, , \label{ADM}
\end{eqnarray}
and so $\sqrt{-\gamma} := \sqrt{-\det{\gamma_{ab}}} = \epsilon N \sqrt{\chi}$
with $\epsilon=+1$ if $N>0$ and $\epsilon=-1$ if $N<0$. Due to the fact
$\tau$ is time-like and $\sigma$ is space-like $- N^2 + \lambda^2 \chi < 0$
and $\chi>0$. Taking into account (\ref{ADM}) $S[\gamma^{ab}, X^{\mu}]$
acquires the form
\begin{eqnarray}
S[X^{\mu}, {\widetilde P}_{\mu}, {\ut \lambda}, \lambda] & = & \int_R d\tau
\int_{\Sigma} d \sigma \left [ {\dot X}^{\mu} {\widetilde P}_{\mu} - \left (
{\ut \lambda} {\widetilde{\widetilde H}} + \lambda {\widetilde D} \right )
\right ]\, , \label{Haction}
\end{eqnarray}
where the dependency of the phase space variables and Lagrange multipliers in
terms of the Lagrangian variables is
\begin{eqnarray}
{\widetilde P}_{\mu}= -\frac{2 \alpha \epsilon \sqrt{\chi}}{N} {\dot X}^{\nu}
g_{\mu\nu} + \frac{2\alpha \epsilon \lambda \sqrt{\chi}}{N} {X'}^{\nu}
g_{\mu\nu} \, , \quad {\ut \lambda} = -\frac{N}{4 \alpha\epsilon\sqrt{\chi}}
\, , \label{DEF}
\end{eqnarray}
with
\begin{eqnarray}
{\widetilde{\widetilde H}} = {\widetilde P}_{\mu} {\widetilde P}_{\nu}
g^{\mu\nu} + 4 \alpha^2 {X'}^{\mu} {X'}^{\nu} g_{\mu\nu} \, , \quad
{\widetilde D} = {X'}^{\mu} {\widetilde P}_{\mu} \, , \label{constII}
\end{eqnarray}
the Hamiltonian and diffeomorphism constraints, respectively. Here
${X'}^{\mu} = \frac{\partial X^{\mu}}{\partial \sigma}$. The standard
variational principle is formed with the action (\ref{Haction}) and with the
boundary conditions
\begin{eqnarray}
X^{\mu}( \tau_i ,\sigma) & = & x^{\mu}_i (\sigma)\, , \quad i=1,2,
\label{SBC}
\end{eqnarray}
where $x^{\mu}_i (\sigma)$ are the initial (at $\tau_1$) and final (at
$\tau_2$) string configurations.

The variation of $S[X^{\mu}, {\widetilde P}_{\mu}, {\ut \lambda}, \lambda]$
under (\ref{SBC}) and the condition $\delta S=0$ yield the equations of
motion
\begin{eqnarray}
{\dot X}^{\mu}  = 2 \ut{\lambda} {\widetilde P}_{\nu} g^{\mu\nu} + \lambda
{X'}^{\mu} \, ,\quad  {\dot {\widetilde P}}_{\mu} = \ut{\lambda}\,
{\widetilde{\widetilde Y}}^{\theta\phi} \frac{\partial
g_{\theta\phi}}{\partial X^{\mu} } + \left ( \ut{\lambda} 8 \alpha^2
{X'}^{\nu} g_{\mu\nu} + \lambda {\widetilde P}_{\mu} \right )'\, ,
\label{EMOTION}
\end{eqnarray}
and the constraints
\begin{eqnarray}
{\widetilde{\widetilde H}} = 0 \, , \quad
{\widetilde D}  = 0\, , \label{Csurface}
\end{eqnarray}
where ${\widetilde{\widetilde Y}}^{\theta\phi}  = {\widetilde P}_{\mu}
{\widetilde P}_{\nu} g^{\theta\mu} g^{\phi\nu} - 4 \alpha^2 {X'}^{\theta}
{X'}^{\phi} $. To compute the algebra of constraints the Hamiltonian and
diffeomorphism constraints are smeared with arbitrary fields
$\ut{\varepsilon}(\tau,\sigma)$ and $\varepsilon (\tau,\sigma)$
\begin{eqnarray}
H(\ut{\varepsilon}) = \int_{\Sigma} d \sigma\,\,
\ut{\varepsilon} {\widetilde{\widetilde
H}}\, ,\quad
D(\varepsilon) = \int_{\Sigma} d \sigma
\,\,\varepsilon {\widetilde D} \, .
\end{eqnarray}
Then a straightforward computation yields the Poisson brackets between
the constraints
\begin{eqnarray}
\{ H(\ut{\varepsilon})\, ,\, H(\ut{\lambda}) \} =  D(\kappa) + \mbox{B.T.}\,
,\quad \{ D(\varepsilon )\, , \, H(\ut{\varepsilon}) \} =  H ({\cal
L}_{\varepsilon} \ut{\varepsilon}) + \mbox{B.T.}\, , \quad \{
D(\varepsilon)\, , \, D(\lambda) \} =  D({\cal L}_{\varepsilon} \lambda ) +
\mbox{B.T.}, \label{algebra}
\end{eqnarray}
where ${\cal L}_{\varepsilon} \lambda =\varepsilon \lambda' - \lambda
\varepsilon'$, ${\cal L}_{\varepsilon} \ut{\varepsilon}= \varepsilon
\ut{\varepsilon}' -\ut{\varepsilon} \varepsilon'$, $\kappa=16
\alpha^2(\ut{\varepsilon}\ut{\lambda}'-\ut{\lambda}\ut{\varepsilon}')$, and
B.T. stands for boundary terms. Therefore, ${\widetilde{\widetilde H}}$ and
${\widetilde D}$ are first class and the theory has $D-2$ physical degrees of
freedom ($2(D-2)$ in the phase space) per space point.

{\it Geometric perspective}. Even though the Hamiltonian and diffeomorphism
constraints have an evident meaning, it is interesting to see what the
constraint surface means from the perspective of the geometry of the induced
metric $h_{ab}$. By using the equation of motion for $X^{\mu}$ and the
definitions of the induced metric components the constraints (\ref{constII})
become
\begin{eqnarray}
{\widetilde{\widetilde H}} = \frac{1}{4 {\ut{\lambda}}^2} h_{\tau\tau}
- \frac{\lambda}{2 {\ut{\lambda}}^2} h_{\tau\sigma} + \left (
\frac{\lambda^2}{4 {\ut{\lambda}}^2} + 4 \alpha^2 \right )
h_{\sigma\sigma}\, , \quad
{\widetilde D}  = \frac{1}{2 \ut{\lambda}} h_{\tau\sigma} -
\frac{\lambda}{2\ut{\lambda}} h_{\sigma\sigma}\, .
\end{eqnarray}
Therefore, the constraint surface (\ref{Csurface}) just means
\begin{eqnarray}
\frac{1}{4 {\ut{\lambda}}^2} h_{\tau\tau}
- \frac{\lambda}{2 {\ut{\lambda}}^2} h_{\tau\sigma} + \left (
\frac{\lambda^2}{4 {\ut{\lambda}}^2} + 4 \alpha^2 \right )
h_{\sigma\sigma} = 0 \, , \quad
\frac{1}{2 \ut{\lambda}} h_{\tau\sigma} -
\frac{\lambda}{2\ut{\lambda}} h_{\sigma\sigma} = 0 \, .
\end{eqnarray}
From the second equation $h_{\tau\sigma}$ can be plugged into the first and
then the constraint surface looks like
\begin{eqnarray}
h_{\tau\tau} = \left ( -16 {\ut{\lambda}}^2 \alpha^2 + \lambda^2 \right )
h_{\sigma\sigma}\, , \quad h_{\tau\sigma} = \lambda h_{\sigma\sigma}\, .
\label{ConstGeo}
\end{eqnarray}
These relationships among the world sheet metric components and the Lagrange
multipliers just came as a consequence of describing string dynamics from a
canonical perspective. Also it is possible to write down the induced metric
$h_{ab}$ in terms of the phase space variables and Lagrange multipliers
\begin{eqnarray}
h_{\tau\tau} & = & {\dot X}^{\mu} {\dot X}^{\nu} g_{\mu\nu} =
4 {\ut{\lambda}}^2 {\widetilde P}_{\mu} {\widetilde P}_{\nu} g^{\mu\nu} +
4 \ut{\lambda} \lambda {\widetilde P}_{\mu} {X'}^{\mu} +
\lambda^2 {X'}^{\mu} {X'}^{\nu} g_{\mu\nu}\, , \nonumber\\
h_{\tau\sigma} & = & {\dot X}^{\mu} {X'}^{\nu} g_{\mu\nu} =
2 \ut{\lambda} {\widetilde P}_{\mu} {X'}^{\mu} +
\lambda {X'}^{\mu} {X'}^{\nu} g_{\mu\nu} \, , \nonumber\\
h_{\sigma\sigma} & = &  {X'}^{\mu} {X'}^{\nu} g_{\mu\nu} \, ,
\end{eqnarray}
where the equation of motion for $X^{\mu}$ was used. These expressions
mean that in order to compute the induced metric on the world sheet is
necessary 1) fix the gauge and thus to determine the Lagrange multipliers
$\ut{\lambda}$, and $\lambda$, 2) with each particular choice for the
Lagrange multipliers solve the equations of motion, 3) plug these solutions
into the constraints and into the gauge fixing conditions to drop the gauge
freedom completely, finally, 4) insert the final expression for $X^{\mu}$
together with the Lagrange multipliers into the RHS of the metric components.

By using (\ref{constII}) the metric components can be written in terms of the
constraints, the Lagrange multipliers and just one metric component
\begin{eqnarray}
h_{\tau\tau} = 4 {\ut{\lambda}}^2 {\widetilde{\widetilde H}} + 4
\ut{\lambda} \lambda {\widetilde D} + \left ( -16 {\ut{\lambda}}^2 \alpha^2
+ \lambda^2 \right ) h_{\sigma\sigma}\, , \quad
h_{\tau\sigma} = 2 \ut{\lambda} {\widetilde D} + \lambda h_{\sigma\sigma}\, .
\end{eqnarray}
Thus, on the constraint surface
\begin{eqnarray}
h_{\tau\tau} = \left ( -16 {\ut{\lambda}}^2 \alpha^2
+ \lambda^2 \right ) h_{\sigma\sigma}\, , \quad
h_{\tau\sigma} = \lambda h_{\sigma\sigma}\, ,
\end{eqnarray}
as expected [see Eq. (\ref{ConstGeo})].

In addition, it is interesting to compute the energy-momentum tensor
(\ref{tensor}) in terms of the phase space variables and the Lagrange
multipliers. By doing this the components of $T=T_{ab} d \xi^a d\xi^b$ become
\begin{eqnarray}
T_{\tau\tau} = -2\alpha {\ut{\lambda}}^2 \left ( 1 + \frac{\lambda^2}{16
\alpha^2 {\ut{\lambda}}^2 } \right ) {\widetilde{\widetilde H}} - 4 \alpha
\lambda \ut{\lambda} {\widetilde D}\, , \quad T_{\tau\sigma} =
-\frac{\lambda}{8\alpha} {\widetilde{\widetilde H}} - 2 \alpha \ut{\lambda}
{\widetilde D} \, , \quad T_{\sigma\sigma} =
-\frac{1}{8\alpha}{\widetilde{\widetilde H}} \, ,
\end{eqnarray}
and thus the energy-momentum tensor vanishes on the constraint surface.
However, the trace of the energy-momentum zero vanishes identically, $T_a\,^a
= T_{ab} g^{ab} =0 {\widetilde{\widetilde H}} +0 {\widetilde D}=0$, and not
just on the constraint surface.

{\it Gauge transformations}. Before computing the gauge transformation on the
phase space variables and Lagrange multipliers induced by the first class
constraints, it is interesting to compute the {\it finite} transformation of
these variables due to both Poincar\'e and Weyl invariance, i.e., it is
assumed in this part of the paper that the background metric $g_{\mu\nu}$ is
the Minkowski one $\eta_{\mu\nu}$.

i) Poincar\'e invariance is ${\cal X}^{\mu} (\tau,\sigma) =
\Lambda^{\mu}\,_{\nu} X^{\nu} (\tau,\sigma) + a^{\mu}$,
${\widetilde{\gamma}}_{ab} (\tau,\sigma) =\gamma_{ab}(\tau,\sigma)$ with
$\Lambda^{\mu}\,_{\nu}$ a Lorentz transformation and $a^{\mu}$ a translation.
Using the explicit form of $\gamma_{ab}$ in (\ref{ADM}) and the definition of
the momentum (\ref{DEF}) {\it finite} Poincar\'e invariance means, in the
Hamiltonian framework, that the phase space variables and Lagrange
multipliers transform as
\begin{eqnarray}
{\cal X}^{\mu}(\tau,\sigma) = \Lambda^{\mu}\,_{\nu} X^{\nu} (\tau,\sigma) +
a^{\mu}\, , \quad {\widetilde{\cal P}}_{\mu} (\tau,\sigma) =
\Lambda_{\mu}\,^{\nu} {\widetilde P}_{\nu}(\tau,\sigma) \, , \quad
\ut{\lambda'} (\tau,\sigma) = \ut{\lambda} (\tau,\sigma) \, ,\quad \lambda'
(\tau,\sigma) = \lambda (\tau,\sigma) \, . \label{Henri}
\end{eqnarray}
ii) Two-dimensional Weyl invariance is ${\cal X}^{\mu} (\tau,\sigma) =
X^{\mu} (\tau,\sigma)$,
${\widetilde{\gamma}}_{ab}(\tau,\sigma)=e^{2\omega(\tau,\sigma)}
\gamma_{ab}(\tau,\sigma)$ for arbitrary $\omega(\tau,\sigma)$. Using the
explicit form of $\gamma_{ab}$ in (\ref{ADM}) and the definition of the
momentum (\ref{DEF}) {\it finite} two-dimensional Weyl invariance means, in
the Hamiltonian framework, that the phase space variables and Lagrange
multipliers transform as
\begin{eqnarray}
{\cal X}^{\mu}(\tau,\sigma) = X^{\mu} (\tau,\sigma) \, , \quad
{\widetilde{\cal P}}_{\mu} (\tau,\sigma) = \frac{1}{\epsilon'} {\widetilde
P}_{\mu} (\tau,\sigma) \, ,\quad \ut{\lambda'} (\tau,\sigma)  = \epsilon'
\ut{\lambda} (\tau,\sigma) \, , \quad \lambda' (\tau,\sigma) =  \lambda
(\tau,\sigma)  \, ; \quad \epsilon'=\pm 1 \, , \label{Weyl}
\end{eqnarray}
due to the fact $\chi$, and $N$ transform as $\widetilde{\chi} (\tau,\sigma)
= e^{2\omega(\tau,\sigma)} \chi (\tau,\sigma)$, $\widetilde{N} (\tau,\sigma)
= \epsilon' e^{\omega(\tau,\sigma)} N (\tau,\sigma)$. Nevertheless, only
$\epsilon'=1$ leaves action ({\ref{Haction}) invariant under the
transformation (\ref{Weyl}).

Let us go back to general case, namely, when the background metric
$g_{\mu\nu}$ is left arbitrary. In this case, it is easy to compute the
infinitesimal gauge transformation induced by the constraints
({\ref{Csurface}) on the phase space variables \cite{Dirac}
\begin{eqnarray}
{\cal X}^{\mu} (\tau,\sigma) & = & X^{\mu} (\tau,\sigma) + \{ X^{\mu}
(\tau,\sigma), H(\ut{\varepsilon}) \} +
\{ X^{\mu} (\tau,\sigma), D(\varepsilon) \}\, , \nonumber\\
& = & X^{\mu} (\tau,\sigma)+ 2 \left ( \ut{\varepsilon} \,\,
g^{\mu\nu} {\widetilde P}_{\nu} \right ) (\tau,\sigma) +
{\cal L}_{\varepsilon} X^{\mu} (\tau,\sigma) \, , \nonumber\\
{\widetilde {\cal P}}_{\mu} (\tau,\sigma) & = & {\widetilde P}_{\mu}
(\tau,\sigma) + \{ {\widetilde P}_{\mu} (\tau,\sigma), H(\ut{\varepsilon}) \}
+ \{ {\widetilde P}_{\mu} (\tau,\sigma), D(\varepsilon) \}
\, , \nonumber\\
& = & {\widetilde P}_{\mu} (\tau,\sigma) + \ut{\varepsilon} \,\,
{\widetilde{\widetilde Y}}^{\theta\phi} \frac{\partial
g_{\theta\phi}}{\partial X^{\mu}} (\tau,\sigma) - 8\alpha^2 \ut{\varepsilon}
(\tau,y) \delta(y,\sigma) {X'}^{\nu}
(\tau,y) g_{\mu\nu} (\tau,y) \mid^{y=\sigma_2}_{y=\sigma_1}  \nonumber\\
& & + (8 \alpha^2 \ut{\varepsilon} {X'}^{\nu} g_{\mu\nu})'(\tau,\sigma)
-\varepsilon(\tau,y) \delta(y,\sigma) {\widetilde P}_{\mu} (\tau,y)
\mid^{y=\sigma_2}_{y=\sigma_1} + {\cal L}_{\varepsilon} {\widetilde P}_{\mu}
(\tau,\sigma)\, , \label{diff}
\end{eqnarray}
${\cal L}_{\varepsilon} X^{\mu}= \varepsilon {X'}^{\mu}$, ${\cal
L}_{\varepsilon} {\widetilde P}_{\mu}= (\varepsilon {\widetilde P}_{\mu})'$.
The gauge symmetry (\ref{diff}) is associated with the two-dimensional
diffeomorphism invariance of the theory. For closed strings no boundary terms
appear in the constraints algebra (\ref{algebra}) and the transformation law
for the phase space variables simplifies accordingly
\begin{eqnarray}
{\cal X}^{\mu} (\tau,\sigma) & = & X^{\mu} (\tau,\sigma)+ 2 \left (
\ut{\varepsilon} \,\,g^{\mu\nu} {\widetilde P}_{\nu} \right ) (\tau,\sigma) +
{\cal L}_{\varepsilon} {X}^{\mu} (\tau,\sigma)
\, , \nonumber\\
{\widetilde {\cal P}}_{\mu} (\tau,\sigma) & = & {\widetilde P}_{\mu}
(\tau,\sigma) + \ut{\varepsilon} \,\, {\widetilde{\widetilde Y}}^{\theta\phi}
\frac{\partial g_{\theta\phi}}{\partial X^{\mu}} (\tau,\sigma) + (8 \alpha^2
\ut{\varepsilon} {X'}^{\nu} g_{\mu\nu})'(\tau,\sigma) + {\cal
L}_{\varepsilon} {\widetilde P}_{\mu}(\tau,\sigma)\, , \label{GaugeII}
\end{eqnarray}
while the Lagrange multipliers transform as
\begin{eqnarray}
\ut{\lambda'} (\tau,\sigma) =  \ut{\lambda}(\tau,\sigma)+
\ut{\dot\varepsilon}+ {\cal L}_{\varepsilon} \ut{\lambda} - {\cal
L}_{\lambda} \ut{\varepsilon} \,\,\, ,\quad \lambda' (\tau,\sigma)  = \lambda
(\tau,\sigma) + {\dot\varepsilon} + {\cal L}_{\varepsilon} \lambda - \kappa\,
. \label{GaugeV}
\end{eqnarray}
Taking into account (\ref{GaugeII}) and (\ref{GaugeV}) the gauge
transformation induces a transformation in the action for the closed string
\begin{eqnarray}
S [{\cal X}^{\mu} , {\widetilde {\cal P}}_{\mu} , \ut{\lambda'}, \lambda' ]
& = & S[X^{\mu} , {\widetilde P}_{\mu} , \ut{\lambda}, \lambda ] +
\int^{\sigma_2}_{\sigma_1} d \sigma \left [ \{ X^{\mu}
(\tau,\sigma) \, , \,  H \} {\widetilde P}_{\mu} (\tau,\sigma)-
\left ( \ut{\varepsilon} {\widetilde{\widetilde H}} +
\varepsilon {\widetilde D} \right ) \right ]^{\tau=\tau_2}_{\tau=\tau_1}\, ,
\end{eqnarray}
with
\begin{eqnarray}
H = \int^{\sigma_2}_{\sigma_1} d\sigma \left (
\ut{\varepsilon} {\widetilde{\widetilde H}} + \varepsilon
{\widetilde D} \right ) = H(\ut{\varepsilon}) + D(\varepsilon)\, .
\end{eqnarray}
After a direct computation
\begin{eqnarray}
S [{\cal X}^{\mu} , {\widetilde {\cal P}}_{\mu} , \ut{\lambda'}, \lambda' ] &
= & S[X^{\mu} , {\widetilde P}_{\mu} , \ut{\lambda}, \lambda ] +
\int^{\sigma_2}_{\sigma_1} d\sigma \left [ \ut{\varepsilon} \left (
{\widetilde P}_{\mu} {\widetilde P}_{\nu} g^{\mu\nu} - 4 \alpha^2 {X'}^{\mu}
{X'}^{\nu} g_{\mu\nu} \right ) \right ]^{\tau =\tau_2}_{\tau=\tau_1}\, ,
\nonumber\\
& = & S[X^{\mu} , {\widetilde P}_{\mu} , \ut{\lambda}, \lambda ] +
\int^{\sigma_2}_{\sigma_1} d\sigma \left ( \ut{\varepsilon}
{\widetilde{\widetilde Y}}^{\mu\nu} g_{\mu\nu} \right)^{\tau
=\tau_2}_{\tau=\tau_1}\, .
 \label{change}
\end{eqnarray}
Therefore $S[X^{\mu} , {\widetilde P}_{\mu} , \ut{\lambda}, \lambda ]$ is not
gauge-invariant and behaves in the same way as the action for the
relativistic free particle [cf Ref. \cite{Mon0501}]. The reason why
$S[X^{\mu} , {\widetilde P}_{\mu} , \ut{\lambda}, \lambda ]$ fails to be
gauge invariant is because the Hamiltonian constraint is quadratic in the
momenta like in systems with finite degrees of freedom
\cite{Teitelboim1982,Henneaux1992}. In spite of this, fully gauge invariant
actions under finite gauge symmetries for systems with finite degrees of
freedom were built in Ref. \cite{Mon0501}. Now, those ideas are here extended
to field theory. In the particular case when the background metric
$g_{\mu\nu}$ is constant, for instance when $g_{\mu\nu}$ is the Minkowski
metric $\eta_{\mu\nu}$, the action for the closed string
\begin{eqnarray}
S_{inv}[X^{\mu} , {\widetilde P}_{\mu} , \ut{\lambda}, \lambda ]& = &
\int^{\tau_2}_{\tau_1} d\tau \int^{\sigma_2}_{\sigma_1} d\sigma \left [ {\dot
X}^{\mu} {\widetilde P}_{\mu} - \left ( {\ut \lambda} {\widetilde{\widetilde
H}} + \lambda {\widetilde D} \right ) \right ] - \frac12
\int^{\sigma_2}_{\sigma_1} d\sigma \left ( X^{\mu} {\widetilde P}_{\mu}
\right )^{\tau=\tau_2}_{\tau=\tau_1}\, , \label{fully}
\end{eqnarray}
is, at first order, fully gauge-invariant.

{\it Observables}. Dirac observables or observables for short are functions
defined on the reduced phase space of the theory. They are constant along the
gauge orbits of the constraint surface and thus they have weakly vanishing
Poisson brackets with the Hamiltonian and diffeomorphism constraints. At
infinitesimal level this means observables must be gauge invariant under the
gauge transformation (\ref{GaugeII}). From the transformation law for the
phase space variables (\ref{GaugeII}) it is clear that for closed strings
propagating in a constant background $g_{\mu\nu}$ the linear and angular
momentum
\begin{eqnarray}
P_{\mu} = \int^{\sigma_2}_{\sigma_1} d\sigma {\widetilde P}_{\mu}
(\tau,\sigma) \, , \quad M^{\mu\nu} =  \int^{\sigma_2}_{\sigma_1} d\sigma
\left ( X^{\mu} (\tau,\sigma) {\widetilde P}^{\nu} (\tau,\sigma) - X^{\nu}
(\tau,\sigma) {\widetilde P}^{\mu} (\tau,\sigma) \right ) \, ,\label{obs}
\end{eqnarray}
respectively, are observables; and thus the values of $P_{\mu}$ and $M^{\mu}$
are independent of any particular choice for the gauge conditions. From their
own definitions apparently the linear momentum $P_{\mu}$ and the angular
momentum $M^{\mu\nu}$ depend on $\tau$. However, by computing their
derivative with respect to $\tau$ and using the equations of motion $ {\dot
P}_{\mu} =0= {\dot M}^{\mu\nu}$. Therefore, $P_{\mu} $ and $M^{\mu\nu}$ are
indeed independent of the time coordinate $\tau$ and, of course, of the space
coordinate $\sigma$. Notice that two string configurations having same
$P_{\mu}$ and $M^{\mu\nu}$ do not represent the same physical string
configuration because $P_{\mu}$ and $M^{\mu\nu}$ do not label the full
reduced phase space of string theory. Moreover, it is possible to build other
observables from combinations of the previous ones as, for example, the
square mass $M^2$ of the closed string $M^2=-P^{\mu} P_{\mu}$. Up to here, it
has been shown $P_{\mu}$ and $M^{\mu\nu}$ are observables because they are
invariant under the gauge transformation generated by the first class
constraints (\ref{GaugeII}). However, what about Poincar\'e and Weyl
invariance? Notice that under Poincar\'e invariance (\ref{Henri}) the linear
momentum $P_{\mu}$ and the angular momentum $M^{\mu\nu}$ transform as ${\cal
P}_{\mu} = \Lambda_{\mu}\,^{\nu} P_{\nu}$, ${M'}^{\mu\nu}=
\Lambda^{\mu}\,_{\alpha} \Lambda^{\nu}\,_{\beta} M^{\alpha\beta} + ( a^{\mu}
\Lambda^{\nu}\,_{\alpha} - a^{\nu} \Lambda^{\mu}\,_{\alpha} ) P^{\alpha}$
while under two-dimensional Weyl invariance (\ref{Weyl}) they are fully
gauge-invariant (taking $\epsilon'=1$) ${\cal P}_{\mu} = P_{\mu}$,
${M'}^{\mu\nu}= M^{\mu\nu}$ because ${\widetilde{\cal P}}_{\mu}
(\tau,\sigma)= {\widetilde P}_{\mu} (\tau,\sigma)$ and ${\cal
X}^{\mu}(\tau,\sigma) = X^{\mu} (\tau,\sigma)$. This will be very important
in a moment.  Due to the fact the ten observables $P_{\mu}$ and $M^{\mu\nu}$
in a Minkowski target are associated with its isometries (Killing vector
fields), it is natural to expect that the analogous of $P_{\mu}$ and
$M^{\mu\nu}$ in arbitrary backgrounds, where Poincar\'e invariance is lost,
would be associated with their Killing vector fields too. In fact, if $v=
v^{\mu} (X) \frac{\partial}{\partial X^{\mu}}$ is a Killing vector field of
the background spacetime $g=g_{\mu\nu} (X) d X^{\mu} d X^{\nu}$, then a
straightforward application of Noether's theorem to (\ref{polyakov}) implies
\begin{eqnarray}
O_v & = & \int^{\sigma_2}_{\sigma_1} d \sigma {\widetilde P}_{\mu}
(\tau,\sigma) v^{\mu}(X(\tau,\sigma))\, ,
\end{eqnarray}
are observables [see \cite{Carter} for an alternative approach in the case of
$p$-branes]. But, what about if backgrounds had no isometries? This simply
would mean that there would be no observables associated with isometries,
however, still there would be observables, i.e., invariant entities under the
transformation (\ref{GaugeII}) associated with the true physical degrees of
freedom of strings.

Other quantities used in string theory are the `center of mass' coordinates
of the string
\begin{eqnarray}
X^{\mu} (\tau) & = & \int^{\sigma_2}_{\sigma_1} d \sigma X^{\mu}
(\tau,\sigma) \, . \label{center}
\end{eqnarray}
However, the `center of mass' coordinates of a closed string are {\it not}
observables under the transformation (\ref{GaugeII}). This might be source of
confusion with intuition. Certainly, $X^{\mu}(\tau)$ are measurable
quantities, but measurable quantities are not, in general, observables of the
theory. In addition, under Poincar\'e invariance (\ref{Henri}) the `center of
mass' coordinates transform as ${\cal X}^{\mu} (\tau) = \Lambda^{\mu}\,_{\nu}
X^{\nu} (\tau) + a^{\mu} (\sigma_2 - \sigma_1)$ while under Weyl invariance
(\ref{Weyl}) they are fully gauge-invariant ${\cal X}^{\mu} (\tau) = X^{\mu}
(\tau)$ because ${\cal X}^{\mu}(\tau,\sigma) = X^{\mu} (\tau,\sigma)$.

So far, it has been exhibited the transformation laws for $P_{\mu}$ and
$M^{\mu\nu}$ under i) Poincar\'e invariance (\ref{Henri}), ii)
two-dimensional Weyl invariance (\ref{Weyl}), and iii) the transformation law
associated with the first class constraints (\ref{GaugeII}).  Let us compare
with gravity. Here, gravity is not string gravity, rather, it is Einstein's
general relativity. In four dimensional general relativity is neither the
symmetry of the kind associated with global Lorentz invariance (\ref{Henri})
nor the symmetry of the kind associated with two-dimensional Weyl invariance
(\ref{Weyl}) [this type of symmetry is also not present in the
Dirac-Nambu-Goto action \cite{Nambu}], rather, the gauge symmetry present in
general relativity is of the same kind that the one coming from the first
class constraints (\ref{constII}), (\ref{GaugeII}). Indeed, from last
computations important notions can been drawn which make shape to the meaning
of observables in generally covariant theories, in particular, for general
relativity. The first lesson from $P_{\mu}$ and $M^{\mu\nu}$ is that they are
gauge-invariant under the gauge symmetry associated with the first class
constraints (\ref{GaugeII}). The second lesson is that $P_{\mu}$ and
$M^{\mu\nu}$ are {\it independent} of the time and space coordinates $\tau$
and $\sigma$; respectively, which label the points on the world sheet.
Therefore, in any generally covariant theory having Hamiltonian and
diffeomorphism constraints, as general relativity, must happen the same
phenomenon: observables must be coordinate independent entities too. In
string theory, on the other hand, fields have physical meaning because they
are attached to the fixed background $\eta_{\mu\nu}$, and thus $P_{\mu}$ (or
$M^{\mu\nu}$) can be measured in any `external' Lorentz reference frame. The
relationship between the values of the linear momentum $P_{\mu}$ measured
from any two `external' Lorentz observers is ${\cal P}_{\mu} =
\Lambda_{\mu}\,^{\nu} P_{\nu}$ with $\Lambda_{\mu}\,^{\nu}$ a matrix in the
Lorentz group. However, the presence of `external' observers placed in the
background manifold is a peculiar fact of string theory and it is not a
general property of generally covariant theories, for instance, in general
relativity `external' observers are not allowed; there is not a background
manifold `outside' of spacetime where `external' observers sit to see how
spacetime propagates, rather, dynamics of the gravitational field must be
described from an `inside' viewpoint. This is a key conceptual difference of
general relativity with respect to string theory. Nevertheless, as already
mentioned it is still true that in general relativity observables must be
coordinate independent entities as well, and this fact implies a major
problem in gravity. In general relativity spacetime coordinates are attached
to `observers' placed in some reference frame, so how can an `observer'
measure some observable, say in his (her) laboratory, if observables are
independent of spacetime coordinates? In other words, `local' observables in
general relativity or in any other generally covariant theory are not allowed
because of diffeomorphism invariance \cite{Carlo}.

{\it Gauge fixing}. In any gauge theory, determinism forces it to identify
gauge related phase space variables as a single point in the reduced phase
space of the gauge theory, and the total number of these orbits span its
physical phase space. At classical level, good gauge conditions help to
single out these physical degrees of freedom because they intersect just once
the gauge orbits on the constraint surface. On the other hand, in quantum
theory there are essentially two ways to proceed: i) reduce then quantize or
ii) quantize then reduce. In i) the relevance of a good gauge fixing is
clear. Standard quantization of strings is of the kind i) and so it is
important to have good gauge conditions to do that. Before going to that
point, some words about other unfortunate choice for the gauge conditions
\begin{eqnarray}
\ut{\lambda} = 0 \,  \quad \lambda=1\, , \label{bad}
\end{eqnarray}
usually found in the literature. Due to the fact $\tau$ is time-like and
$\sigma$ is space-like $\gamma_{\tau\tau}= - N^2 + \lambda^2 \chi <0$ and
$\gamma_{\sigma\sigma}= \chi>0$ must hold, which means $\lambda^2 < 16
\alpha^2 {\ut{\lambda}}^2$. It is clear (\ref{bad}) does not satisfy this
condition. Putting it in a different manner, the choice (\ref{bad}) breaks
down the causal structure on the world sheet because with such a choice
$\tau$ becomes space-like and $\sigma$ becomes time-like.

To fix consistently the gauge degrees of freedom in the action
(\ref{polyakov}), the components of the inverse of the world sheet metric
$\gamma^{ab}$ will be considered as dynamical variables. In this case, there
are three additional constraints, since the canonical momenta associated to
$\gamma^{ab}$ are weakly equal to zero
\begin{equation}
{\widetilde\pi}_{ab} \approx 0\, , \quad a,b=1,2 \, .
\end{equation}
In this approach, instead of (\ref{Haction}), the canonical action is
\begin{eqnarray}
S[X^{\mu}, \gamma^{ab}, {\widetilde P}_{\mu}, {\widetilde\pi}_{ab}, {\ut
\lambda_1}, \lambda_1, \lambda^{ab}] & = & \int_R d\tau \int_{\Sigma} d
\sigma \left [ {\dot X}^{\mu} {\widetilde P}_{\mu} +\dot \gamma^{ab}
{\widetilde\pi}_{ab} - \left ( \lambda^{ab} {\widetilde\pi}_{ab} + {\ut
\lambda_1} {\widetilde{\widetilde H}} + \lambda_1 {\widetilde D} \right )
\right ]\, . \label{Hactionpi}
\end{eqnarray}
This action becomes the action (\ref{Haction}) when Weyl invariant variables
are used, all the constraints being first class. So, to fix the gauge, five
gauge conditions are needed. Notice that the Lagrange multipliers $({\ut
\lambda_1}, \lambda_1)$ are not exactly the same Lagrange multipliers of
(\ref{Haction}). Both sets are related by
\begin{eqnarray}
\lambda_1 =\lambda + \rho\, , \quad \ut{\lambda_1}= \ut{\lambda} +
\ut{\rho}\, ,
\end{eqnarray}
where the additional arbitrary parts $(\rho, \ut{\rho})$ appear from the fact
that the constraints (\ref{constII}) are secondary ones in this approach.

Now, the conformal fixing of the world sheet metric will be considered
\begin{equation}\label{gaugecon}
\gamma^{\tau\tau}=-1, \ \ \ \gamma^{\tau\sigma}=0, \ \ \
\gamma^{\sigma\sigma}=1\, .
\end{equation}
These gauge conditions set the Lagrange multipliers $\lambda^{ab}=0$,
however, from the infinitesimal gauge transformation for the intrinsic metric
\begin{equation}
\delta \gamma_{ab} = \eta^c \partial_c \gamma_{ab} + \partial_a \eta^c
\gamma_{cb} + \partial_b \eta^c \gamma_{ca} + 2 \omega \gamma_{ab}\, ,
\end{equation}
it follows that the conditions (\ref{gaugecon}) do {\it not} fix completely
the gauge freedom of the gauge parameters $(\eta^a , \omega)$. These
parameters are only restricted to satisfy the differential equations
\begin{equation}
\ddot \eta^\tau - \eta^{\tau\prime\prime} =0,\ \ \ddot \eta^\sigma -
\eta^{\sigma\prime\prime} =0, \ \
 \dot \eta^\tau = \eta^{\sigma\prime} = -\omega\, . \label{Lageq}
\end{equation}
The remaining gauge freedom is associated to the conformal group in two
dimensions, which is infinite-dimensional. The Lagrangian gauge parameters
$\eta^a$ and the Hamiltonian ones $(\ut{\varepsilon},\varepsilon)$ are
related by
\begin{eqnarray}
\eta^{\tau} = \frac{\ut{\varepsilon}}{\ut{\lambda_1}}\, , \quad
\eta^{\sigma}= - \frac{\ut{\varepsilon}}{\ut{\lambda_1}} \lambda_1 +
\varepsilon \, . \label{Lagpa}
\end{eqnarray}
Now to fix the additional gauge freedom associated with the constraints
(\ref{constII}), the light-cone gauge conditions are chosen (in the case of
the closed string propagating in the Minkowski spacetime)
\begin{equation}
\tau = A X^+/ p^+ \, , \quad \quad P^+ = \frac{B}{2} p^+ \, ,
\end{equation}
$A$, $B$ constants, and where the light-cone coordinates are given by
\begin{equation}
X^{\pm} = {1 \over \sqrt{2}} (X^0 \pm X^{D-1})\, , \quad P^{\pm} = {1 \over
\sqrt{2}} ({\widetilde P}^0 \pm {\widetilde P}^{D-1}) \, .
\end{equation}
These gauge conditions allow it to fix the Lagrange multiplier ${\ut
\lambda_1}=\frac{1}{AB}$, however, they do {\it not} fix completely
$\lambda_1$, rather, it is left as an arbitrary $\tau$-dependent function,
$\lambda_1= \lambda_1 (\tau)$. In addition, $\ut{\varepsilon}=0$ and
$\varepsilon=\varepsilon (\tau)$. By plugging $\ut{\varepsilon}=0$ and
$\varepsilon=\varepsilon(\tau)$ into (\ref{Lagpa}) $\eta^{\tau}=0$,
$\eta^{\sigma}=\varepsilon(\tau)$. By inserting in (\ref{Lageq}), those
equations set $\omega=0$, and $\eta^{\sigma}=\varepsilon(\tau) =a_1 \tau
+a_2$, with $a_1$, $a_2$ constants. So, the system is still invariant under
the $\tau$-dependent coordinate transformations
\begin{equation}
\tau' = \tau \, , \quad \sigma' = \sigma + f(\tau)\, .
\end{equation}
This residual gauge invariance is important in the quantum theory of the
string \cite{Henneaux2}.

\section{Relatives of bosonic string theory}

\subsection{Pure diffeomorphism bosonic string theory}
The algebra of constraints for string theory allows it to define a new theory
that looks like the action for string theory except that it has no
Hamiltonian constraint, being its dynamics attached to the diffeomorphism
constraint only. This theory is defined by
\begin{eqnarray}
S[X^{\mu}, {\widetilde P}_{\mu}, \lambda] & = & \int^{\tau_2}_{\tau_1} d\tau
\int^{\sigma_2}_{\sigma_1} d\sigma \left [ {\dot X}^{\mu} {\widetilde
P}_{\mu} - \lambda {\widetilde D} \right ]\, , \label{HK}
\end{eqnarray}
with ${\widetilde D} = {X'}^{\mu} {\widetilde P}_{\mu}$, ${X'}^{\mu} =
\frac{\partial X^{\mu}}{\partial \sigma}$. The algebra of constraints closes
and thus ${\widetilde D}$ is first class. The equations of motion are
\begin{eqnarray}
{\dot X}^{\mu} = \lambda {X'}^{\mu} \, , \quad {\dot{\widetilde P}}_{\mu}
 = (\lambda {\dot {\widetilde P}}_{\mu})' \, .
\end{eqnarray}
The theory defined by (\ref{HK}) contains string theory (\ref{Haction}) as a
sub-sector of its space of solutions because the theory defined by
(\ref{HK}}) has one more physical degree of freedom than string theory
(\ref{Haction}). This is a general fact, always that a diffeomorphism
constraint appears in the formalism of generally covariant theories it closes
with itself, and thus it is possible to drop some of the other constraints
involved in their algebra and thus to build larger theories which will
contain the formers as sub-sectors, like the one defined by (\ref{HK}) which
emerged from (\ref{actionII}). It is pretty obvious that a similar
construction holds for the bosonic $p$-branes where instead of having one
single diffeomorphism constraint there will be a finite number of them.
However, at first sight, a supersymmetric version of (\ref{HK}) might not be
allowed. A more radical interpretation for the theory defined by (\ref{HK})
and its relationship with string theory is to see (\ref{HK}) as a kind of
$M$-theory, and to consider different sectors of this $M$-theory as ones
defined by different Hamiltonian constraints ${\widetilde{\widetilde H}}$'s.
Notice that in the case when the spatial surface is closed, the action
(\ref{HK}) is fully gauge-invariant under the gauge transformation generated
by ${\widetilde D}$. Due to the fact ${\widetilde{\widetilde H}}$ is missing
in (\ref{HK}), a deep analysis of (\ref{HK}) can help to understand better
the role that the Hamiltonian constraint ${\widetilde{\widetilde H}}$ plays
in string theory both classical and quantum mechanically. Finally, it is
important to mention that the theory defined by (\ref{HK}) plays the same
role with respect to string theory (\ref{Haction}) as the Husain-Kuchar model
plays with respect to self-dual gravity for self-dual gravity is a sub-sector
of the space of solution of the Husain-Kuchar model \cite{Husain}. Actually,
to have a better analogy it would be desirable to have a Lagrangian form for
(\ref{HK}).

\subsection{Tensionless bosonic string theory with constraints linear in
the momenta}
String action (\ref{Haction}) has another relative in the case
when the background metric $g_{\mu\nu}$ is constant, say the Minkowski or
Euclidean metric $\eta_{\mu\nu}$. The later is defined by setting $\alpha=0$
in the constraints, namely, it is defined by the action
\begin{eqnarray}
S[X^{\mu}, {\widetilde P}_{\mu}, {\ut \lambda}, \lambda] & = &
\int^{\tau_2}_{\tau_1} d\tau \int^{\sigma_2}_{\sigma_1} d\sigma
\left [ {\dot X}^{\mu} {\widetilde P}_{\mu} - \left ( {\ut \lambda}
{\widetilde{\widetilde H}} + \lambda {\widetilde D} \right ) \right ]
\, , \label{actionII}
\end{eqnarray}
with
\begin{eqnarray}
{\widetilde{\widetilde H}} = {\widetilde P}_{\mu} {\widetilde P}_{\nu}
\eta^{\mu\nu} \, ,\quad {\widetilde D} = {X'}^{\mu} {\widetilde P}_{\mu} \, ,
\label{constIV}
\end{eqnarray}
where ${X'}^{\mu} = \frac{\partial X^{\mu}}{\partial \sigma}$. Obviously,
this action can not be obtained from the Polyakov action (\ref{polyakov})
because if $\alpha$ were equal zero then the RHS of (\ref{polyakov}) would
vanish too.

Let us focus in the case when the spatial slice of the `world sheet' is
closed. The algebra of constraints is
\begin{eqnarray}
\{ H(\ut{\varepsilon})\, ,\, H(\ut{\lambda}) \} = 0 \, ,\quad \{
D(\varepsilon )\, , \, H(\ut{\varepsilon}) \} = H ({\cal L}_{\varepsilon}
\ut{\varepsilon}) \, , \quad \{ D(\varepsilon)\, , \, D(\lambda) \} = D({\cal
L}_{\varepsilon} \lambda )\, .
\end{eqnarray}
Under the gauge symmetry generated by the constraints, the action changes,
according to (\ref{change}), as
\begin{eqnarray}
S [{\cal X}^{\mu} , {\widetilde {\cal P}}_{\mu} , \ut{\lambda'}, \lambda' ]
& = & S[X^{\mu} , {\widetilde P}_{\mu} , \ut{\lambda}, \lambda ] +
\int^{\sigma_2}_{\sigma_1} d\sigma \left (
\ut{\varepsilon} {\widetilde{\widetilde H}}
\right )^{\tau=\tau_2}_{\tau=\tau_1}\, . \label{changeII}
\end{eqnarray}
Therefore, the boundary term is proportional to the Hamiltonian constraint,
and thus the action is gauge invariant on the constraint surface. A similar
situation appears in general relativity expressed in terms of Ashtekar
variables \cite{Mon0401}. In Ref. \cite{Mon0401} it was no built the fully
gauge-invariant action associated with the self-dual action, however, this
could be carried out.

Let us come back to the action (\ref{actionII}) and construct, following the
steps of \cite{Mon0501}, its fully gauge-invariant action. This action is
given by
\begin{eqnarray}
S_{inv}[X^{\mu} , {\widetilde P}_{\mu}, \ut{\lambda}, \lambda ]& = &
\int^{\tau_2}_{\tau_1} d\tau \int^{\sigma_2}_{\sigma_1} d\sigma \left [ {\dot
X}^{\mu} {\widetilde P}_{\mu} - \left ( {\ut \lambda} {\widetilde{\widetilde
H}} + \lambda {\widetilde D} \right ) \right ] - \frac12
\int^{\sigma_2}_{\sigma_1} d\sigma \left ( X^{\mu} {\widetilde P}_{\mu}
\right )^{\tau=\tau_2}_{\tau=\tau_1} \, . \label{actionIII}
\end{eqnarray}
A straightforward computation shows that, at first order, $S_{inv}[X^{\mu} ,
{\widetilde P}_{\mu} , \ut{\lambda}, \lambda ]$ is fully gauge-invariant. The
boundary term in (\ref{actionIII}) induces the canonical transformation
\begin{eqnarray}
q^0 & = & \frac{1}{2} \ln{\left ( \frac{X^0}{{\widetilde P}_0}\right )} \quad
, \quad p_0 =
X^0 {\widetilde P}_0 \, , \nonumber\\
q^1& = & \frac{1}{2} \ln{\left ( \frac{X^1}{{\widetilde P}_1}\right )} \quad
, \quad p_1 =
X^1 {\widetilde P}_1 \, , \nonumber\\
... & = & .....\nonumber\\
q^D & = & \frac{1}{2} \ln{\left ( \frac{X^D}{{\widetilde P}_D} \right )}
\quad , \quad p_D = X^D {\widetilde P}_D\, , \quad (\mbox{no sum over $D$})\,
.
\end{eqnarray}
In terms of the new phase space variables $S_{inv}$ reads
\begin{eqnarray}
S_{inv} [q^{\mu} , p_{\mu} , \ut{\lambda}, \lambda ] & = &
\int^{\tau_2}_{\tau_1} d\tau \int^{\sigma_2}_{\sigma_1} d\sigma \left [
{\dot q}^{\mu} p_{\mu} - \left ( \ut{\lambda} {\widetilde {\widetilde H}} +
\lambda {\widetilde D} \right )\right ]\, ,
\end{eqnarray}
with
\begin{eqnarray}
{\widetilde{\widetilde H}} = p_{\mu} e^{-2q^{\nu}} \eta^{\mu\nu}
\, ,\quad
{\widetilde D} = \frac12 \left ({p'}_{\mu} l^{\mu} + 2 p_{\mu} {q'}^{\mu}
\right )\, ,
\end{eqnarray}
with $l^{\mu} = (1,1,...,1)$ and it was assumed a diagonal background metric
$\eta_{\mu\nu}$. Notice that the constraints are linear and homogeneous in
the momenta (and in their derivatives).


\end{document}